\documentclass[12pt]{article}
\usepackage{epsfig,amsmath,amsfonts,amssymb,amstext,afterpage,psfrag,
}
\usepackage{slashed}
\usepackage{color}
\usepackage[square,comma,sort&compress,numbers]{natbib}

\allowdisplaybreaks
\usepackage{hyperref}

\newcommand{\lp}{\left(}
\newcommand{\rp}{\right)}

\newcommand{\al}{\alpha}
\newcommand{\be}{\beta}

\newdimen\TW
\usepackage{color}
\definecolor{light-gray}{gray}{0.9}
\definecolor{dark-gray}{gray}{0.7}
\usepackage{tikz}

\long\def\symbolfootnote[#1]#2{\begingroup%
\def\thefootnote{\fnsymbol{footnote}}\footnote[#1]{#2}\endgroup}

\def\spose#1{\hbox to 0pt{#1\hss}}
\def\lsim{\mathrel{\spose{\lower 3pt\hbox{$\mathchar"218$}}
 \raise 2.0pt\hbox{$\mathchar"13C$}}}
\def\gsim{\mathrel{\spose{\lower 3pt\hbox{$\mathchar"218$}}
 \raise 2.0pt\hbox{$\mathchar"13E$}}}

\begin{document}

\begin{titlepage}

\begin{flushright}
{\small
LMU-ASC~40/23\\ 
December 2023
}
\end{flushright}

\vspace{0.5cm}
\begin{center}
{\Large\bf \boldmath                                               
  Two-Higgs Doublet Model Matched to\\
\vspace*{0.3cm}                                                            
  Nonlinear Effective Theory
\unboldmath}
\end{center}

\vspace{0.5cm}
\begin{center}
  {\sc G. Buchalla, F. K\"onig, Ch. M\"uller-Salditt and F. Pandler} 
\end{center}

\vspace*{0.4cm}

\begin{center}
Ludwig-Maximilians-Universit\"at M\"unchen, Fakult\"at f\"ur Physik,\\
Arnold Sommerfeld Center for Theoretical Physics, 
D--80333 M\"unchen, Germany
\end{center}

\vspace{1.5cm}
\begin{abstract}
\vspace{0.2cm}\noindent
We use functional methods to match the Two-Higgs Doublet Model
with heavy scalars in the nondecoupling regime to the appropriate
nonlinear effective field theory, which takes the form of an
electroweak chiral Lagrangian (HEFT). The effective Lagrangian
is derived to leading order in the chiral counting.
This includes the loop induced $h\to\gamma\gamma$ and
$h\to Z\gamma$ local terms, which enter at the same chiral order
as their counterparts in the Standard Model.
An algorithm is presented that allows us to compute the coefficient
functions to all orders in $h$.
Some of the all-orders results are given in closed form.
The parameter regimes for de\-coup\-ling, nondecoupling and alignment
scenarios in the effective field theory
context and some phenomenological implications are briefly
discussed.
\end{abstract}

\vfill

\end{titlepage}

\section{Introduction}
\label{sec:intro}

Indirect effects of New Physics (NP) at colliders can be consistently
described with effective field theories (EFTs), where the new heavy
particles are integrated out.
Applying this approach to electroweak symmetry
breaking and Higgs-boson properties, the nonlinear EFT in the
form of an electroweak chiral Lagrangian (EWChL, also refered
to as nonlinear Higgs-sector EFT, or HEFT)
\cite{Feruglio:1992wf,Bagger:1993zf,Koulovassilopoulos:1993pw,Burgess:1999ha,Wang:2006im,Grinstein:2007iv,Contino:2010mh,Contino:2010rs,Buchalla:2012qq,Alonso:2012px,Alonso:2012pz,Buchalla:2013rka,Delgado:2014jda,Buchalla:2015qju,Alonso:2016oah,Cohen:2020xca,Gomez-Ambrosio:2022qsi,Gomez-Ambrosio:2022why,Delgado:2023ynh}
provides us with the most natural framework~\cite{Buchalla:2015wfa}.
It is economic and general and properly accounts for nondecoupling
effects in the scalar sector.
While the EFT is model independent, matching its parameters to a specific
scenario connects the EFT coefficients to a given UV theory.
Recently, there has been renewed interest in the Two-Higgs Doublet
Model (2HDM) \cite{Branco:2011iw} and the treatment of its
properties at the electroweak scale in an EFT approximation
\cite{Dmytriiev:2022asw,Banta:2023prj,Dawson:2023ebe,Arco:2023sac}
(for earlier work see e.g. \cite{Ciafaloni:1996ur}).
Our motivation for addressing this topic is essentially twofold.
First, we would like to investigate the description
of the 2HDM in the nondecoupling regime, which corresponds to
interesting regions of parameter space.
Second, our analysis exemplifies the structure of the
Higgs-EWChL in the context of the 2HDM as a prototypical extension
of the Higgs sector.
In addition, we use functional methods throughout, which make the
calculations rather efficient and transparent.
Exploiting the advantages of the functional approach, we
go beyond the existing literature in computing higher terms in the
Higgs functions, including some all-orders results in powers
of the Higgs field $h$, and an algorithmic prescription for their general
derivation.
The method used in the present study has been developed in detail
in \cite{Buchalla:2016bse}, where it was applied to the matching of a
singlet extension of the SM to the nonlinear EFT.

The paper is organized as follows.
In Section 2 we introduce polar coordinates for the scalar sector of
the 2HDM, which are especially convenient for the matching to the
nonlinear EFT. In Section 3 we perform the matching of the 2HDM in the
nondecoupling regime to the leading-order (LO) chiral Lagrangian at tree level,
integrating out the heavy scalars by functional methods.
The matching calculation is extended to the one-loop induced
$h\to\gamma\gamma$ and $h\to Z\gamma$ local EFT operators in
Section 4. Section 5 summarizes important aspects of the 2HDM
parameter space with heavy-scalar masses (of order TeV),
including the decoupling, nondecoupling and alignment regimes.
Some phenomenological implications are discussed in Section 6,
before we conclude in Section 7.
An Appendix contains the solution $H_0(h)$ of the LO equations of motion
(eom) for the heavy scalar field $H_0$ to all orders in $h$
(Appendix A), the one-loop matching for $h\to\gamma\gamma$ and
$h\to Z\gamma$ to all orders in $h$ (Appendix B), and explicit expressions
for the parameters of the 2HDM scalar potential (Appendix C).

\section{2HDM scalar sector in polar coordinates}
\label{sec:2hdmu}

The scalar sector of the 2HDM consists of two complex doublets
$H_1$, $H_2$, both in the fundamental representation of the weak
gauge group $SU(2)$ and with weak hypercharge ${\cal Y}=1/2$.
It is convenient to define the conjugate doublets
$\tilde H_n\equiv i\sigma_2 H^*_n$, with $n=1,2$, and the matrix fields
\begin{align}\label{sndef}
S_n\equiv (\tilde H_n, H_n)
\end{align}
The Lagrangian of the scalar sector can then be expressed as
\begin{align}\label{lhsn}
{\cal L}_S = \frac{1}{2} \langle D_\mu S^\dagger_n D^\mu S_n\rangle - V 
\end{align}
where $\langle\ldots\rangle$ denotes the trace, a sum over $n$ is understood,
and $V$ is the potential to be discussed below.
Following~\cite{Dittmaier:2022ivi}, the matrix fields $S_n$ can be
written in polar coordinates as 
\begin{align}\label{surdef}
S_n\equiv U R_n\, , \qquad R_n=
\frac{1}{\sqrt{2}}\Bigl[(v_n + h_n)\boldsymbol{1} + i C_n \sigma_a \rho_a\Bigr]
\end{align}
Here $\sigma_a \equiv 2 T_a$, $a=1,2,3$, are the Pauli matrices, 
and $U\equiv\exp(2i\varphi_a T_a/v)$ is the matrix of the electroweak
Goldstone bosons, where $v=246\, {\rm GeV}$ is the electroweak
vacuum expectation value (vev).
The~vevs of the two Higgs doublet fields are $v_1$ and $v_2$,
respectively, with $v^2_1 + v^2_2 = v^2$, and
\begin{align}\label{c12v12}
C_1 =-\frac{v_2}{v}\equiv - \sin\beta\, , \qquad
C_2 =\frac{v_1}{v}\equiv  \cos\beta  
\end{align}

Using the decomposition in (\ref{surdef}), the eight real
degrees of freedom in the complex doublets $H_n$ are
expressed through the eight real fields $\varphi_a$, $\rho_a$ and $h_n$.
The electroweak quantum numbers of $S_n$ and $U$ imply that the covariant
derivative reads
\begin{align}\label{covdsu}
  D_\mu \Phi = \partial_\mu \Phi + i g W_\mu \Phi - i  g' B_\mu \Phi T_3
  \qquad {\rm for} \quad \Phi=S_n , \, U 
\end{align}
where $W^\mu=W^\mu_a T_a$ and $B^\mu$ are the gauge fields
of $SU(2)_L$ and $U(1)_Y$. It follows from (\ref{surdef}) that
\begin{align}\label{covdr}
D_\mu R_n = \partial_\mu R_n + i g' B_\mu [T_3, R_n]
\end{align}
Consequently, $h_{1,2}$ and $\rho_3$ are electroweak singlets,
whereas $\rho_{1,2}$ are singlets of $SU(2)_L$, but charged under $U(1)_Y$.
Hence, 
\begin{align}\label{covdrhoa}
D_\mu h_n = \partial_\mu h_n \qquad {\rm and}\qquad
D_\mu \rho_a = \partial_\mu \rho_a +  g' B_\mu \, \varepsilon_{ab3} \, \rho_b
\end{align}
For $a=1,2$, this can also be written in terms of the eigenstates
$\rho^\pm$ of charge and hypercharge (with $Q={\cal Y}=\pm 1$) as
\begin{align}\label{covdrhopm}
D_\mu \rho^\pm = \partial_\mu \rho^\pm \pm i  g' B_\mu \rho^\pm\, ,\qquad
\rho^\pm =\frac{1}{\sqrt{2}}(\rho_1 \mp i \rho_2)
\end{align}
Inserting (\ref{surdef}) into (\ref{lhsn}), the kinetic term
becomes
\begin{align}\label{lskin}
{\cal L}_{S,kin} &= \frac{1}{2} \langle D_\mu S^\dagger_n D^\mu S_n\rangle = 
\nonumber\\
&= \frac{1}{4}\langle D_\mu U^\dagger D^\mu U\rangle
\Bigl[ (v_n + h_n)^2 + \rho_a \rho_a\Bigr] +
\frac{1}{2}\partial_\mu h_n\partial^\mu h_n +\frac{1}{2} D_\mu\rho_a D^\mu\rho_a
\nonumber\\
&+ \langle i U^\dagger D_\mu U T_a\rangle
     \Bigl[ \varepsilon_{abc}\, \rho_b D^\mu\rho_c +
       C_n(\rho_a\partial^\mu h_n - D^\mu\rho_a h_n)\Bigr]                   
\end{align}

The potential in (\ref{lhsn}) can be written as \cite{Dittmaier:2022ivi}
\begin{align}\label{vs1s2}
  V= &\frac{m^2_{11}}{2} \langle S^\dagger_1 S_1\rangle +  
       \frac{m^2_{22}}{2} \langle S^\dagger_2 S_2\rangle -
       m^2_{12} \langle S^\dagger_1 S_2\rangle \nonumber\\
     &+\frac{\lambda_1}{8} \langle S^\dagger_1 S_1\rangle^2
       +\frac{\lambda_2}{8} \langle S^\dagger_2 S_2\rangle^2 +
 \frac{\lambda_3}{4}\langle S^\dagger_1 S_1\rangle\langle S^\dagger_2 S_2\rangle
  \nonumber\\
 &+\lambda_4 \langle S^\dagger_1 S_2P_+\rangle  \langle S^\dagger_1 S_2P_-\rangle
   +\frac{\lambda_5}{2}\Bigl(  \langle S^\dagger_1 S_2 P_+\rangle^2
       + \langle S^\dagger_1 S_2 P_- \rangle^2 \Bigr)
\end{align} 
in terms of the matrix fields $S_n$ from (\ref{sndef}).
$P_\pm =(1\pm\sigma_3)/2$ are projection operators.
Here we assume invariance under $S_1\to - S_1$, $S_2\to S_2$, softly broken
by $m^2_{12}$, and CP invariance, so that all parameters
in (\ref{vs1s2}) are real.

When $S_n$ is expressed as in (\ref{surdef}), the Goldstone field $U$
disappears from the potential $V$ in (\ref{vs1s2}), which becomes a function
of $h_n$ and $\rho_a$. The vevs $v_{1,2}$ are defined such that
terms linear in $h_{1,2}$ vanish.
The terms quadratic in the fields are diagonalised by
$\rho^\pm$, $\rho_3$, and by $h$ and $H$, which are related to
$h_{1,2}$ by
\begin{equation}\label{h1h2al}
  \left(  \begin{array}{c} h_1 \\ h_2  \end{array} \right) =
  \left(\begin{array}{cc}
          c_\alpha & - s_\alpha \\ s_\alpha & c_\alpha  \end{array} \right)
\left(  \begin{array}{c} H \\ h  \end{array} \right)
\end{equation}  
Here and in the following, we define $\cos\phi\equiv c_\phi$ and
$\sin\phi\equiv s_\phi$ for generic angles $\phi$.
The mass eigenstates of the scalar sector are then
given by $h$, identified as the observed Higgs at $m_h=125\,{\rm GeV}$,
and the additional scalars $H\equiv H_0$, $H^\pm\equiv\pm i\rho^\pm$
and $A_0\equiv -\rho_3$.

The eight parameters of $V$ in (\ref{vs1s2}),
$m^2_{11}$, $m^2_{22}$, $m^2_{12}$, $\lambda_1,\ldots, \lambda_5$,
can be traded for the vevs, the particle masses, the
Higgs mixing angle and the soft breaking term:
\begin{align}\label{physpar1}
  v_1,\quad v_2,\quad m_h,\quad M_0\equiv M_{H_0},
  \quad M_H\equiv M_{H^\pm}, \quad M_A\equiv M_{A_0},
  \quad s_\alpha, \quad m^2_{12}   
\end{align}  
or, equivalently,
\begin{align}\label{physpar2}
  v,\quad \tan\beta\equiv t_\beta = v_2/v_1,\quad m_h,\quad M_0,
  \quad M_H, \quad M_A, \quad c_{\beta-\alpha},
  \quad   \overline{m}^2 \equiv\frac{m_{12}^2}{s_{\beta} c_{\beta}}  
\end{align}

Dropping an irrelevant additive constant, the potential
finally takes the form
\begin{align}\label{vdizi}
  V = &\frac{1}{2} m^2_h h^2 + \frac{1}{2} M^2_0 H^2
        + M^2_H H^+ H^- + \frac{1}{2} M^2_A A^2_0 \nonumber\\
      &- d_1 h^3 - d_2 h^2 H -d_3 h H^2 - d_4 H^3 
      -d_5 h H^+ H^- - d_6 h A^2_0 - d_7 H H^+ H^- - d_8 H A^2_0 \nonumber\\
      &-z_1 h^4 - z_2 h^3 H - z_3 h^2 H^2 - z_4 h H^3 - z_5 H^4 \nonumber\\
      &-z_6 h^2 H^+ H^- - z_7 h H H^+ H^- -z_8 H^2 H^+ H^-
        - z_9 (H^+ H^-)^2 -z_{10}h^2 A^2_0 \nonumber\\
     &-z_{11} h H A^2_0 -z_{12} H^2 A^2_0 - z_{13} H^+ H^- A^2_0 - z_{14} A^4_0 
\end{align}  
The coefficients can be found in Appendix \ref{sec:potpar}.

The scalar sector couples to fermions through Yukawa interactions.
We assume a type-II Yukawa sector given by the
Lagrangian \cite{Branco:2011iw}
\begin{align}\label{yukawatypeII}
  \mathcal{L}_Y =
  -\bar q_L H_1 Y_{d} d_R - \bar q_L \tilde{H}_2 Y_{u} u_R
  -\bar\ell_L H_1 Y_{e} e_R +\text{h.c.}
\end{align}
where $q_L=(u_L,d_L)^T$ and $\ell_L=(\nu_L,e_L)^T$ are the left-handed
doublets, and $u_R$, $d_R$, $e_R$ the right-handed singlets. The latter
may be collected into $q_R=(u_R,d_R)^T$ and $\ell_R=(\nu_R,e_R)^T$. We
suppress generation indices, which are understood for the fermion
fields in (\ref{yukawatypeII}).
$\mathcal{L}_Y$ can be written in terms of the matrix fields $S_n$
in (\ref{surdef}) as
\begin{align}\label{yukawasn}
  \mathcal{L}_Y =
  -\bar q_L Y_{d} S_1 P_- q_R - \bar q_L Y_{u} S_2 P_+ q_R
  -\bar\ell_L Y_{e} S_1 P_- \ell_R +\text{h.c.}
\end{align}

\section{Tree-level matching in the nondecoupling regime}
\label{sec:treemat}

The full Lagrangian of the 2HDM can be written as
\begin{align}\label{l2hdm}
\mathcal{L}_{2HDM} = \mathcal{L}_0 + \mathcal{L}_{S,kin} - V + \mathcal{L}_Y  
\end{align}
where the scalar sector is represented by $\mathcal{L}_{S,kin}$,
$V$ and $\mathcal{L}_Y$  from (\ref{lskin}), (\ref{vdizi}) and
(\ref{yukawasn}), and $\mathcal{L}_0$ denotes the unbroken
Standard Model (SM),
\begin{align}\label{lsm0}
\mathcal{L}_0 = &-\frac{1}{2} \langle G_{\mu\nu}G^{\mu\nu}\rangle
-\frac{1}{2}\langle W_{\mu\nu}W^{\mu\nu}\rangle 
-\frac{1}{4} B_{\mu\nu}B^{\mu\nu}\nonumber\\
&+\bar q_L i\!\not\!\! Dq_L +\bar\ell_L i\!\not\!\! D\ell_L
 +\bar u_R i\!\not\!\! Du_R +\bar d_R i\!\not\!\! Dd_R
    +\bar e_R i\!\not\!\! De_R 
\end{align}

In terms of the model parameters in (\ref{physpar1}), (\ref{physpar2})
the nondecoupling limit is defined by the hierarchy
\begin{align}\label{defnondec}
v\sim m_h\sim \overline{m} \ll M_0\, ,M_H\, , M_A \sim M_S
\end{align}
with $t_\beta$ and $c_{\beta-\alpha}$ of order unity in general.
To leading order, all terms in the effective Lagrangian that are unsuppressed
by the heavy scale $M_S$ (of order $(M_S)^0$) have to be retained.

The procedure of integrating out the heavy scalars at tree level in the
nondecoupling scenario has been described in detail
in~\cite{Buchalla:2016bse}. It consists of the following steps:
\begin{itemize}
\item
  The equation of motion (eom) is solved to obtain the heavy field $H_0(h)$
  to LO in the heavy-mass limit, ${\cal O}(M^0_S)$.
  This requires the LO terms in the full-theory Lagrangian
  of order $M^2_S$. A closed-form solution for $H_0(h)$ is derived
  in Appendix~\ref{sec:exh0}.
  The ${\cal O}(M^2_S)$-Lagrangian contains the heavy fields $A_0$ and
  $H^\pm$ only at quadratic order or higher. Contributions with
  only internal lines from these fields, therefore, cannot arise
  at tree level. Integrating them out at tree level and to LO
  then implies $A_0=H^\pm=0$.
\item
  The eom solutions $H_0=H_0(h)$ and $A_0=H^\pm=0$ are inserted into the
  Lagrangian (\ref{l2hdm}). The ${\cal O}(M^2_S)$-terms cancel
  and an expression of ${\cal O}(M^0_S)$ in the heavy-mass expansion
  is obtained.
\item
  The field redefinition
\begin{align}\label{htilde}
\tilde h=\int^h_0 \sqrt{1+\left(\frac{dH_0(s)}{ds}\right)^2}\, ds
\end{align}
is performed to achieve a canonically normalized kinetic term
for the Higgs field $\tilde h$.
For notational convenience we will drop the tilde in the end,
taking $\tilde h\to h$.
\end{itemize}

Proceeding in this way, the effective theory takes the form
of an electroweak chiral Lagrangian at chiral dimension two,
${\cal L}={\cal L}_0 + {\cal L}_{Uh,2}$, with
\begin{align}\label{luhlo}
\mathcal{L}_{Uh,2} =& \frac{v^2}{4} \langle D_{\mu} U^{\dag} D^{\mu} U \rangle 
\left( 1 + F_U(h)\right) + \frac{1}{2} \partial_{\mu} h \partial^{\mu} h - V(h) 
\nonumber \\ 
&- \Bigl[ \bar q_L \left( {\cal M}_u + \sum_{n=1}^{\infty}  {\cal M}_{u}^{(n)} 
\left( \frac{h}{v} \right)^n  \right) U P_+ q_R   + \bar q_L    
\left( {\cal M}_d + \sum_{n=1}^{\infty} {\cal M}_{d}^{(n)} 
\left(  \frac{h}{v} \right)^n  \right) U P_- q_R  \nonumber \\
& + \bar \ell_L \left( {\cal M}_e + \sum_{n=1}^{\infty} {\cal M}_e^{(n)} 
\left( \frac{h}{v}  \right)^n \right) U P_- \ell_R + \mathrm{h.c.} \Bigr]
\end{align}
where
\begin{align}\label{fuh}
  F_U(h) = 2 s_{\beta-\alpha}\frac{h}{v} +
  \left(1- \frac{s_{2\alpha}}{s_{2\beta}} c^2_{\beta-\alpha} \right)
  \left(\frac{h}{v}\right)^2  -\frac{4}{3}
  \frac{s^2_{2\alpha}}{s^2_{2\beta}} c^2_{\beta-\alpha} s_{\beta-\alpha}
  \left(\frac{h}{v}\right)^3 +\ldots
\end{align}

\begin{align}\label{vvh}
  V(h) &= \frac{v^2 m_h^2}{2} \bigg\{\bigg(\frac{h}{v}\bigg)^2 +
  \bigg[s_{\beta-\alpha} + \frac{2 c_{\beta-\alpha}^2 c_{\beta+\alpha}}{s_{2\beta}}
 \bigg(1 -\frac{\overline{m}^2}{m_h^2} \bigg) \bigg]\bigg(\frac{h}{v}\bigg)^3 
   \nonumber\\
 &+ \bigg[\frac{1}{4} - \frac{c_{\beta-\alpha}^2}{4 s_{2\beta}^2}
 \bigg(\frac{1}{6} (7 - 12 c_{2(\beta+\alpha)} - 19 c_{4\alpha}) -
   (1 - 2 c_{2\alpha} c_{2\beta} - 3 c_{4\alpha}) \frac{\overline{m}^2}{m_h^2}
   \bigg) \bigg] \bigg(\frac{h}{v}\bigg)^4   \nonumber\\
   &- \frac{c_{\beta-\alpha}^2 s_{2\alpha}^2}{2 s_{2\beta}^3}
     \bigg[c_{\beta+\alpha} + 3 c_{\beta-3\alpha} - (2 c_{\beta+\alpha} +
     3 c_{3\beta - \alpha} + 11 c_{\beta-3\alpha}) \frac{\overline{m}^2}{4 m_h^2}
     \bigg] \bigg(\frac{h}{v}\bigg)^5 + \dots  \bigg\}
\end{align}

\begin{align}
 & {\cal M}_u + \sum_{n=1}^{\infty}  {\cal M}_{u}^{(n)}
   \left( \frac{h}{v} \right)^n = \nonumber\\
  & {\cal M}_u  \left[ 1 +
  \frac{c_\alpha}{s_\beta}\frac{h}{v}
  -\frac{c_{\beta-\alpha}}{2}\frac{s^2_\alpha c_\alpha}{s^2_\beta c_\beta}
  \left(\frac{h}{v}\right)^2  -\frac{c_{\beta-\alpha}}{6}
  \frac{s^2_{2\alpha}}{s^2_{2\beta}}(2s_{2\alpha}  -
  (1-2 c_{2\alpha}) t^{-1}_\beta ) \left(\frac{h}{v}\right)^3 +\ldots  \right] \\
  & {\cal M}_d + \sum_{n=1}^{\infty}  {\cal M}_{d}^{(n)}
   \left( \frac{h}{v} \right)^n = \nonumber\\
  & {\cal M}_d  \left[ 1 -
  \frac{s_\alpha}{c_\beta}\frac{h}{v}
  -\frac{c_{\beta-\alpha}}{2}\frac{s_\alpha c^2_\alpha}{s_\beta c^2_\beta}
  \left(\frac{h}{v}\right)^2 + \frac{c_{\beta-\alpha}}{6}
  \frac{s^2_{2\alpha}}{s^2_{2\beta}}(2s_{2\alpha} -
  (1+ 2 c_{2\alpha}) t_\beta ) \left(\frac{h}{v}\right)^3 +\ldots  \right] 
\label{mudh}
\end{align}
The mass matrices ${\cal M}_q$ are related to the Yukawa matrices
in (\ref{yukawasn}) through
\begin{align}\label{myrel}
  {\cal M}_u =\frac{v}{\sqrt{2}} Y_u s_\beta\, ,\qquad
  {\cal M}_d =\frac{v}{\sqrt{2}} Y_d c_\beta
\end{align}  
The expressions for the charged leptons, ${\cal M}_e$,
${\cal M}^{(n)}_e$, are similar to those for the down-quark case. 

Our method reproduces the results of \cite{Arco:2023sac} and gives several
new expressions, the cubic coefficient of $F_U$, the coefficient
of $h^5$ in $V(h)$ and the fermionic couplings.
More generally, the procedure summarized at the beginning of
Section~\ref{sec:treemat}, together with the all-orders expression
for $H_0(h)$ in Appendix~\ref{sec:exh0}, defines an algorithm
to extend the tree-level matching to all orders in $h$.

\subsection*{Other Yukawa interactions}

Besides the type-II Yukawa interactions discussed above, there are
three other possibilities without tree-level flavour changing neutral
currents. Conventionally, these are given by
\begin{itemize}
    \item Type-I
        \begin{equation}                                              
          \mathcal{L} = - \Bar{q}_L Y_d S_2 P_- q_R -
          \Bar{q}_L Y_u S_2 P_+ q_R - \Bar{\ell}_L Y_e S_2 P_- \ell_R
        \end{equation}
    \item Type-X (lepton-specific)
        \begin{equation}                                                  
          \mathcal{L} = - \Bar{q}_L Y_d S_2 P_- q_R -
          \Bar{q}_L Y_u S_2 P_+ q_R - \Bar{\ell}_L Y_e S_1 P_- \ell_R
        \end{equation}
    \item Type-Y (flipped)
        \begin{equation}                                                       
          \mathcal{L} = - \Bar{q}_L Y_d S_1 P_- q_R -
          \Bar{q}_L Y_u S_2 P_+ q_R - \Bar{\ell}_L Y_e S_2 P_- \ell_R
        \end{equation}
\end{itemize}
Using the results of the matching for the type-II 2HDM, it is straightforward
to find the matching for the other Yukawa structures. For example,
in the type-I model, all terms depend only on $S_2$, so the matching
will have the same form as the up-type terms of the type-II 2HDM.
As a result, we find for the type-I 2HDM
    \begin{align}                                                        
      &\mathcal{M}_{u,d,e} + \sum_{n=1}^\infty \mathcal{M}_{u,d,e}^{(n)}
        \bigg(\frac{h}{v}\bigg)^n = \nonumber \\                              
      &\mathcal{M}_{u,d,e} \bigg[1 + \frac{c_\alpha}{s_\beta} \frac{h}{v} -
        \frac{c_{\beta-\alpha}}{2} \frac{s_\alpha^2 c_\alpha}{s_\beta^2 c_\beta}
    \bigg( \frac{h}{v} \bigg)^2 - \frac{c_{\beta-\alpha}}{6}
  \frac{s_{2\alpha}^2}{s_{2\beta}^2} (2 s_{2\alpha} - (1-2 c_{2\alpha}) t_\beta^{-1})
        \bigg(\frac{h}{v} \bigg)^3 + \dots \bigg]                               
    \end{align}
with $\mathcal{M}_{u,d,e} = v Y_{u,d,e} s_\beta/\sqrt{2}$.
It is straightforward to obtain similar expressions for the type-X
and type-Y models.

\section{Nondecoupling effects at one loop}
\label{sec:oneloop}

The procedure of integrating out the heavy scalars can be extended
to one loop using functional methods \cite{Fuentes-Martin:2016uol}.
The most important effects at this order are the local operators
inducing $h\to\gamma\gamma$ and $h\to Z\gamma$ transitions,
because they are loop suppressed in the SM.
The EFT corrections are then at the same loop order as the
leading contributions.
In the 2HDM, the contributions to $h\to\gamma\gamma$ and $h\to Z\gamma$
due to the heavy sector come from charged scalars $H^\pm$ within the loop,
or equivalently from the real fields $\rho_{1,2}$ in~(\ref{surdef}).
To obtain the one-loop contributions with internal $\rho_i$ from
functional integration, the Lagrangian in (\ref{lhsn}) has to be expanded
to quadratic order in these fields.
The quadratic piece takes the form
\begin{align}\label{lrho12}
  {\cal L}^{(2)}_{\rho_{1,2}} = \frac{1}{2} \rho_i \Delta_{ij} \rho_j\, ,\qquad
  \Delta = -D^2 - M^2_H - \hat Y
\end{align}  
where
\begin{align}\label{dxyij}
  D^\mu_{ij} = \partial^\mu \, \delta_{ij} + X^\mu \varepsilon_{ij}
  \equiv (\partial^\mu +\hat X^\mu)_{ij}\, ,\qquad
  \hat Y_{ij} = Y\, \delta_{ij}  
\end{align} 
with $i,j \in \{1,2\}$, $\varepsilon_{ij}$ the two-dimensional
Levi-Civita symbol, and
\begin{align}\label{xazy}
  X^\mu &= e A^\mu +\frac{g}{2c_W} (1-2 s^2_W) Z^\mu \nonumber\\
  -Y &= d_5 h + d_7 H + z_6 h^2 + z_7 h H + z_8 H^2
\end{align} 
Here $e$ is the electromagnetic coupling, $s_W=\sin\theta_W$,
$c_W=\cos\theta_W$ with the Weinberg angle $\theta_W$,
$A$ the photon and $Z$ the $Z$-boson field.
Performing the Gaussian integration of
$\exp(i\int d^4x\, {\cal L}^{(2)}_{\rho_{12}})$ over $\rho_i$
gives the effective Lagrangian \cite{Fuentes-Martin:2016uol}
\begin{align}\label{leffp}
  {\cal L}_{\rm eff}=-\frac{i}{2}\sum_{n=1}^{\infty} \frac{1}{n}
  \int\frac{d^4 p}{(2\pi)^4}\, \Big\langle
  \left(\frac{2i p\cdot D + D^2 + \hat Y}{p^2-M^2_H}\right)^n \Big\rangle
\end{align}

In a weakly-coupled model of the heavy sector, a generic matrix $\hat Y$
scales at most with the first power of $M_H$.
The series in (\ref{leffp}) will then converge and only a finite
number of terms will contribute to any given order in the
$1/M_H$ expansion.
By contrast, in the present nondecoupling scenario we have
$\hat Y\sim M^2_H$, which is of the same order as the denominator
$p^2-M^2_H$. Therefore, an infinite number of terms in the sum over $n$
contributes at a given order in the $1/M_H$ expansion.
However, higher powers of $\hat Y$ come with
higher powers of $h$, since $\hat Y^n={\cal O}(h^n)$ in the field expansion.
As a consequence, the infinite series generates a Higgs-function
$F_O(h)$ that accompanies an EFT operator $O$, as it is characteristic 
for the Higgs-electroweak chiral Lagrangian.
At any given order in $h^n$, the operator coefficient is well-defined and
calculable.

Following this reasoning, we can extract the terms of interest here
from (\ref{leffp}). These contain two factors of the field strength
$\hat X_{\mu\nu}$,
\begin{align}\label{xdd}
  \hat X^{\mu\nu}_{ij}\equiv\left[ D^\mu, D^\nu\right]_{ij}
  =X^{\mu\nu} \varepsilon_{ij}
\end{align}  
corresponding to four covariant derivatives $D$,
along with powers of $\hat Y$.
Neglecting contributions with three or more powers of $h$, we need
to include terms of order $\hat Y$ and $\hat Y^2$.
The result is given by
\begin{align}\label{leffy2}
  32\pi^2\,   {\cal L}_{\rm eff}=
  -\frac{1}{12M^2_H}\langle\hat Y \hat X_{\mu\nu} \hat X^{\mu\nu} \rangle
 +\frac{1}{40 M^4_H}\langle\hat Y^2 \hat X_{\mu\nu} \hat X^{\mu\nu} \rangle
  +\frac{1}{60 M^4_H}\langle (\hat Y \hat X_{\mu\nu})^2 \rangle
\end{align}
which simplifies to
\begin{align}\label{leffysim}
  {\cal L}_{\rm eff}\equiv {\cal L}_{X,4}
  = \frac{X_{\mu\nu} X^{\mu\nu}}{192\pi^2} 
  \,\left[ \frac{Y}{M^2_H} - \frac{Y^2}{2M^4_H} + {\cal O}(h^3)  \right]
\end{align}
Using (\ref{xazy}) and eliminating $H$ in favour of $h$, we obtain
\begin{align}
  {\cal L}_{X,4} &= \frac{e^2}{16\pi^2} \left( A_{\mu\nu} A^{\mu\nu}
  +\frac{1-2s^2_W}{s_Wc_W} A_{\mu\nu} Z^{\mu\nu}\right)\, F_X(h) \label{lhff}\\
  F_X(h) &= \frac{s_{\beta-\alpha}}{6}\frac{h}{v} -\frac{1}{12}
  \left(s^2_{\beta-\alpha}+\frac{s_{2\alpha}}{s_{2\beta}}c^2_{\beta-\alpha}\right)
 \left(\frac{h}{v}\right)^2  + {\cal O}(h^3)  \label{fhfz}
\end{align}
with $A_{\mu\nu}=\partial_\mu A_{\nu}-\partial_\nu A_\mu$,
$Z_{\mu\nu}=\partial_\mu Z_{\nu}-\partial_\nu Z_\mu$.
We note that the field redefinition of $h$, needed to make its kinetic term
canonically normalized, plays no role for $F_X$ through order $h^2$.

The first term $\sim h$ in $F_X(h)$ agrees with the result
of~\cite{Arco:2023sac}, the term $\sim h^2$ is new.
Employing the procedure described above, it is straightforward
to extend the calculation of $F_X$ to higher orders in $h$.
As discussed in Appendix \ref{sec:ynxx}, in the alignment limit
$c_{\beta-\alpha}=0$ the function $F_X$, to all orders in $h$,
takes the simple form
\begin{align}\label{fxfull}
F_X(h) =\frac{1}{6} \ln\left( 1+ \frac{h}{v}\right)
\end{align}
corresponding to the well-known low-energy theorems \cite{Shifman:1979eb}.


\subsection*{Custodial symmetry breaking}

The scalar potential in (\ref{vs1s2}) contains the custodial-symmetry
violating term \cite{Branco:2011iw}
\begin{align}
\Delta V_{CSB}=(\lambda_5 - \lambda_4) \langle S^\dagger_1 S_2 T_3\rangle^2 
\nonumber  
\end{align}
When integrating out the heavy scalars, this term generates the
two-derivative operator
\begin{align}
{\cal L}_{\beta_1} = \beta_1 v^2 \langle U^\dagger D_\mu U T_3\rangle^2 
\nonumber  
\end{align}
but only at the one-loop level.
The coefficient is directly related to the parameter $T$ of
oblique electroweak corrections, $\beta_1=\alpha T/2$, with
$\alpha$ the fine structure constant.
One finds, up to a factor of order
unity, that  \cite{Branco:2011iw}
\begin{align}
\beta_1\sim \frac{\lambda_4 - \lambda_5}{16\pi^2}
        \sim \frac{M^2_A - M^2_H}{16\pi^2 v^2}
\nonumber  
\end{align}
In accordance with the phenomenological requirement of approximate
custodial symmetry, ${\cal L}_{\beta_1}$ cannot be a leading-order effect.
Therefore, the difference $\lambda_4 - \lambda_5$ must be a weak coupling of
${\cal O}(1)$ and carries chiral dimension 2.
${\cal L}_{\beta_1}$ is then counted as a next-to-leading-order (NLO) term
of chiral dimension 4, consistent with $\beta_1$ being small
as a loop factor $1/16\pi^2$ \cite{Buchalla:2013rka}.
The general analysis of such NLO effects is beyond the scope
of the present paper.


\section{Parameter space and the decoupling limit}
\label{sec:decoup}

For the construction of a low-energy EFT, we consider the phenomenologically
viable scenario where the masses of the new scalar degrees of freedom in
the 2HDM are taken to be much larger than the electroweak scale, i.e.
\begin{equation}
   M_S \sim M_0, M_H, M_A \gg m_h \sim v
\end{equation}
Depending on the numerical values of the parameters, we can discern two basic
scenarios, corresponding to weak and strong coupling, respectively.
They are given by
\begin{itemize}
\item \textit{Nondecoupling regime}\footnotemark\, (strong coupling,
  nonlinear EFT) \footnotetext{Although not stated explicitly, this limit
  was used to derive nondecoupling effects in \cite{Arco:2023sac}.}
  \begin{equation}
    1 \ll |\lambda_i| \lesssim 16 \pi^2, \quad m_h \sim v \sim \overline{m}
    \ll M_S \quad \Longrightarrow \quad c_{\beta-\alpha}= \mathcal{O}(1) 
  \end{equation}
While $c_{\beta-\alpha}$ is a priori unconstrained in this regime, we will
also consider the case $c_{\beta-\alpha} \ll 1$, referred to as the
\textit{nondecoupling regime with (quasi-)alignment}. We also note that
the model with $\overline{m}=0$, the $\mathbb{Z}_2$ symmetric 2HDM
without soft breaking, has no decoupling
limit~\cite{Gunion:2002zf,Arhrib:2003vip,Nebot:2019qvr,Faro:2020qyp}.
\item \textit{Decoupling regime}\footnotemark\, (weak coupling, linear EFT)
 \footnotetext{This limit has been studied extensively in \cite{Gunion:2002zf}.
   The model we consider in this work is simpler because of an additional,
   softly broken, $\mathbb{Z}_2$ symmetry $S_1 \rightarrow - S_1$.}
  \begin{equation}
    \lambda_i = \mathcal{O}(1), \quad m_h \sim v \ll \overline{m} \sim M_S
    \quad \Longrightarrow \quad c_{\beta-\alpha} \ll 1
  \end{equation}
\end{itemize}
In the strong-coupling case, we require the $\lambda_i$ to be somewhat below
the nominal strong-coupling limit $M_S \approx 4 \pi v$ corresponding to
$|\lambda_i| \approx 16 \pi^2$. Otherwise a description of the heavy
scalar dynamics in terms of resonances would no longer be valid. To be more
precise, the magnitude of the couplings is constrained by perturbative
unitarity \cite{Huffel:1980sk,Casalbuoni:1987cz,Maalampi:1991fb,Kanemura:1993hm,Akeroyd:2000wc,Horejsi:2005da,Goodsell:2018tti}. For loop corrections to the
constraints, see~\cite{Cacchio:2016qyh,Grinstein:2015rtl,Bahl:2022jnx}.
Generally speaking, these give much stronger bounds,
namely $|\lambda_i| \lesssim 4\pi$. Furthermore, the couplings are
constrained such that the potential is bounded from below and that the
symmetry breaking vacuum is the global minimum of the potential. For the
2HDM with (softly broken) $\mathbb{Z}_2$ symmetry, the necessary and
sufficient conditions on the couplings read \cite{Deshpande:1977rw,Klimenko:1984qx,Maniatis:2006fs,Ivanov:2006yq,Ivanov:2015nea}
\begin{equation}
  \lambda_1 \geq 0 \, ,\quad \lambda_2 \geq 0 \, , \quad \lambda_3
  \geq - \sqrt{\lambda_1 \lambda_2} \, ,
  \quad \lambda_3 + \lambda_4 - |\lambda_5| \geq -\sqrt{\lambda_1 \lambda_2} 
\end{equation}
To satisfy these bounds, the absolute values of the couplings have to be
taken large uniformly, which limits the possible mass splitting between
the heavy scalars. Especially the perturbative unitarity constraints
severely restrict the possible parameter space of the nondecoupling regime.
Nevertheless, masses of $M_S \lesssim 1$ TeV are still possible for
$\overline{m} \sim v$, which clearly fulfills the power counting of the
nondecoupling scenario.

In the decoupling regime, all NP effects are suppressed by powers
of the heavy mass scale $M_S$ as formalised by the Appelquist-Carazzone
decoupling theorem \cite{Appelquist:1974tg}. Several EFT matching
calculations have been performed in the decoupling limit,
see e.g. \cite{Dawson:2023ebe,Perez:1995dc,Dawson:2022cmu}.
A decoupling regime automatically implies the alignment limit
$c_{\beta-\alpha} = 0$, where the $h$-couplings approach their SM
values~\cite{Gunion:2002zf}. An explicit calculation gives
\begin{equation}
  c_{\beta-\alpha}^2 =\frac{v^4}{16 \overline{m}^4} s_{2\beta}^2
  \big[\lambda_1 -\lambda_2 + c_{2\beta} (\lambda_1 + \lambda_2
  - 2 \lambda_{345})\big]^2 + \mathcal{O}(v^6/\overline{m}^6)
\end{equation}
with $\lambda_{345} \equiv \lambda_3 + \lambda_4 + \lambda_5$.
When $\overline{m} \gg v$, this indeed approaches zero. As mentioned above,
there is no similar relation in the nondecoupling regime, and thus,
$c_{\beta-\alpha}$ is unconstrained a priori.

To illustrate the two regimes discussed above, we take the
$hH_0^2$-coupling $d_3$, given in Appendix \ref{sec:potpar}, as an example.
In the nondecoupling regime, $M_0 \sim M_S \gg m_h, \overline{m}$, so
$d_3 = \mathcal{O}(M_S^2)$, whereas in the decoupling regime, the masses and
parameters of the model scale as 
\begin{equation}
  M_0^2, \, M^2_H, \, M^2_A = \overline{m}^2 + \mathcal{O}(v^2) \, ,
  \qquad m_h^2 = \mathcal{O}(v^2) \, ,
  \qquad c_{\beta-\alpha} = \mathcal{O}(v^2/\overline{m}^2) 
\end{equation}
leading to 
\begin{equation}
  d_3 = - \frac{m_h^2}{2v} - v s_\beta^2 c_\beta^2 (\lambda_1 + \lambda_2
  - 2\lambda_{345}) + \mathcal{O}(v^3/\overline{m}^2)
\end{equation}
Evidently, all heavy mass dependence has cancelled. Similar calculations
show that this cancellation works for all $d_i$ and $z_i$. It is now easy
to see that all nondecoupling effects vanish in the decoupling-regime.
Obviously, all tree-level nondecoupling effects vanish in the decoupling
limit, since they are all proportional to $c_{\beta-\alpha}$.
Also the anomalous $h\gamma\gamma$- and $hZ\gamma$-couplings disappear as the
ratios $d_i/M^2_S$ and $z_i/M_S^2$ go to zero in the limit $M_S \to \infty$.

\section{Phenomenological considerations}
\label{sec:pheno}

The simplest way to confront the nondecoupling effects of the 2HDM with
experiment is by using a global HEFT fit. Such a fit has been performed
using LHC run 1 and 2 data \cite{deBlas:2018tjm}, where the authors fit
the couplings of the HEFT Lagrangian in the form
\begin{align}
  \mathcal{L}_{\text{fit}} &= 2 c_V (m_W W_\mu^+ W^{-\mu} +
                             \tfrac{1}{2} m_Z^2 Z_\mu Z^\mu) \frac{h}{v} -
     \sum_\psi c_\psi m_\psi \Bar{\psi} \psi \frac{h}{v} \nonumber \\
  &+ \frac{e^2}{16 \pi^2} c_{\gamma} A_{\mu\nu} A^{\mu\nu} \frac{h}{v}
    + \frac{e^2}{16 \pi^2} c_{\gamma Z} A_{\mu\nu} Z^{\mu\nu} \frac{h}{v} +
    \frac{g_s^2}{16 \pi^2} c_{g}
    \langle G_{\mu\nu} G^{\mu\nu} \rangle \frac{h}{v} 
\end{align}
with $\psi \in \{t, b, c, \tau, \mu\}$\footnotemark \, . Our matching
results are given in Table \ref{tab:MatchingResults}.
\begin{table}[t]
\centering
\begin{tabular}{| c | c || c | c |}
 \hline
 \multicolumn{2}{| c ||}{Tree Level} & \multicolumn{2}{c |}{Loop Level} \\
 \hline\hline
  $c_V$ & $s_{\beta-\alpha}$ & $c_{\gamma}$ & $\frac{s_{\beta-\alpha}}{6}$ \\
  $c_u$ & $s_{\beta-\alpha} + c_{\beta-\alpha} t^{-1}_\beta$ & $c_{\gamma Z}$
 & $\frac{1-2s_W^2}{s_W c_W} \frac{s_{\beta-\alpha}}{6}$\\
  $c_d$ & $s_{\beta-\alpha} - c_{\beta-\alpha} t_\beta $ & $c_{g}$ & $0$ \\
 \hline 
\end{tabular}
\caption{LO matching results for the 2HDM. $c_u$ is the same for
all up-type quarks ($u, c, t$) and $c_d$ is the same for all down-type
quarks ($d, s, b$) and charged leptons ($e, \mu, \tau$).}
\label{tab:MatchingResults}
\end{table}
The strongest
constraint is derived from the Higgs--vector boson coupling~\footnotetext{The
couplings to the lighter fermions are so small that they are not included
in the fit.}
\begin{equation}
c_V = 1.01 \pm 0.06\quad\Longrightarrow\quad s_{\beta-\alpha} \gtrsim 0.95 
\end{equation}
where the given error corresponds to the $68\%$ probability interval.
This motivates the (quasi-)alignment limit as it constrains
$c_{\beta-\alpha} \ll 1$.

Applying the above bound to the anomalous Higgs--photon coupling, we find
\begin{equation}
  c_{\gamma} \in [0.16, 0.17]
\end{equation}
This coupling is particularly important, as it is bounded from below in the
alignment limit. We see here that this is a direct consequence of the bound
on the Higgs--vector boson coupling. From the global HEFT fit, the bound
on $c_{\gamma}$ is given by
\begin{equation}
  c_{\gamma} = 0.05 \pm 0.20 
\end{equation}
which is consistent with the matching prediction. Nevertheless, with more
data from the LHC, it is plausible that the limits on $c_{\gamma}$ could
be sufficiently improved to exclude the nondecoupling regime experimentally.
Local couplings with more than one Higgs in (\ref{lhff}) and (\ref{fhfz}),
such as $h^2 A_{\mu\nu}A^{\mu\nu}$, could in principle
be probed at a photon collider \cite{Asner:2002aa} in a process like
$\gamma\gamma\to hh$.

Aside from using an EFT approach, there is a large amount of literature using
global fits for the 2HDM directly. Depending on the structure of the Yukawa
interactions, these fits can give much stronger bounds on $s_{\beta-\alpha}$
than the global HEFT fit
(see e.g. \cite{Chowdhury:2017aav,Arco:2022xum}).
However, a detailed analysis lies beyond the scope of this work.

\section{Conclusions}
\label{sec:concl}

We presented a systematic derivation of the EFT at the
electroweak scale for the 2HDM in the nondecoupling regime.
In this regime, the EFT takes the form of an electroweak
chiral Lagrangian (nonlinear EFT). Our discussion follows
closely the detailed discussion given in \cite{Buchalla:2016bse}
for the nondecoupling regime of the SM extension with a heavy
scalar singlet.
The scalar sector of the 2HDM is written in polar coordinates,
with a nonlinear representation of the Goldstone fields, which
facilitates the use of functional methods that we employ throughout.
An algorithmic procedure is given, by which the LO EFT Lagrangian
can be worked out to arbitrary order in the Higgs field $h$.
We confirm previous results for the EFT Higgs couplings
and extend the derivation to additional terms.
The main results are displayed in (\ref{luhlo}) -- (\ref{mudh})
and (\ref{lhff}), (\ref{fhfz}). Some all-orders expressions
are given in closed form (Appendix A and B).
We derive the LO EFT Lagrangian, including the fermionic
Yukawa interactions and the loop-induced local terms
for $h\to\gamma\gamma$ and $h\to Z\gamma$. As already pointed out
in \cite{Arco:2023sac,Bhattacharyya:2014oka}, the latter terms
have interesting nondecoupling effects that survive in the
alignment limit. Those are still compatible with present data.
They could be discovered or ruled out
in future measurements of anomalous Higgs-boson couplings.

\section*{Note added}

While this paper was being finalized, the article~\cite{Dawson:2023oce}
appeared on arXiv. It also addresses the HEFT matching of models with
extended scalar sectors and partially overlaps with our results
on the 2HDM.

\section*{Acknowledgements}

This work is supported in part by the Deutsche
Forschungsgemeinschaft (DFG, German Research Foundation)
under Germany’s Excellence Strategy – EXC-2094 – 390783311.

\appendix

\numberwithin{equation}{section}

\section{\boldmath Exact solution for $H_0(h)$}
\label{sec:exh0}

The LO term for $H$ is calculated from the equations of motion
at $\mathcal{O}(M^2_S)$. In particular, we can set $H^\pm = A = 0$ in this
approximation. As a result, retaining only $\mathcal{O}(M^2_S)$ terms the
Lagrangian simplifies to
\begin{equation}\label{eq:HeavyL}
  \mathcal{L}_{M} = - m_{11}^2 \phi_1^2 - m_{22}^2 \phi_2^2 -
  \frac{\lambda_1}{2} \phi_1^4 - \frac{\lambda_2}{2} \phi_2^4 -
   \lambda_{345} \phi_1^2 \phi_2^2  
\end{equation}
where $\phi_n^2 \equiv (v_n + h_n)^2/2$ and
\begin{gather}
     \lambda_1 = \frac{M^2_0}{v^2} \frac{c_\alpha^2}{c_\beta^2} \, , \qquad
     \lambda_2 = \frac{M^2_0}{v^2} \frac{s_\alpha^2}{s_\beta^2} \, , \qquad 
     \lambda_{345}\equiv \lambda_3 + \lambda_4 + \lambda_5 =
     \frac{M^2_0}{v^2} \frac{s_\alpha c_\alpha}{s_\beta c_\beta} \nonumber \\
     m_{11}^2 = - \frac{M^2_0}{2} \bigg(c_\alpha^2 +
     \frac{s_\alpha c_\alpha s_\beta}{c_\beta} \bigg) \, , \qquad
     m_{22}^2 = - \frac{M^2_0}{2} \bigg(s_\alpha^2 +
     \frac{s_\alpha c_\alpha c_\beta}{s_\beta} \bigg) 
\end{gather}
After expressing $h_1$ and $h_2$ through $h$ and $H$,
the two fields take the form
\begin{align}
 \phi_1 &= \frac{1}{\sqrt{2}} (c_\beta v + c_\alpha H - s_\alpha h) \nonumber \\
     \phi_2 &= \frac{1}{\sqrt{2}} (s_\beta v + s_\alpha H + c_\alpha h) 
\end{align}
By defining the combination
\begin{equation}
  R^2 = \frac{c_\alpha}{c_\beta} \phi_1^2 + \frac{s_\alpha}{s_\beta} \phi_2^2 
\end{equation}
the Lagrangian \eqref{eq:HeavyL} can be rewritten as
\begin{equation}\label{lagmr}
  \mathcal{L}_{M} = \frac{M^2_0}{2} (s_\alpha s_\beta + c_\alpha c_\beta) R^2 -
  \frac{M^2_0}{2 v^2} R^4
\end{equation}
At this point we can solve the eom by analogy to the heavy
singlet model studied in \cite{Buchalla:2016bse}.
By direct comparison to the results in the Appendix of
\cite{Buchalla:2016bse}, we can identify
\begin{equation}
  \phi_1 \rightarrow S \, , \quad \phi_2 \rightarrow \phi \, , \quad
  v \rightarrow f \, , \quad s_\beta \rightarrow \sqrt{\xi} \, , \quad
  s_\alpha \rightarrow \sqrt{\omega} \, , \quad M_0 \rightarrow M 
\end{equation}
which exactly reproduces the corresponding terms of the heavy
singlet model. The solution to the eom is then given by
\begin{equation}\label{h0exact}
  H_0(h) = \frac{v + \Big(\frac{s_\alpha^2 c_\alpha}{s_\beta} -
    \frac{s_\alpha c_\alpha^2}{c_\beta} \Big) h}{\frac{s_\alpha^3}{s_\beta} +
    \frac{c_\alpha^3}{c_\beta}} \left[\sqrt{1 -
      \frac{\Big(\frac{s_\alpha^3}{s_\beta} +
        \frac{c_\alpha^3}{c_\beta} \Big)
        \Big(\frac{s_\alpha c_\alpha^2}{s_\beta} +
        \frac{s_\alpha^2 c_\alpha}{c_\beta}
        \Big)h^2}{\Big(v + \Big(\frac{s_\alpha^2 c_\alpha}{s_\beta} -
        \frac{s_\alpha c_\alpha^2}{c_\beta} \Big) h \Big)^2}} - 1\right]
 \end{equation}
 This expression fulfills
\begin{equation}
   R^2 = \frac{v^2}{2} (s_\alpha s_\beta + c_\alpha c_\beta)
   = \frac{v^2}{2} c_{\beta-\alpha}
 \end{equation}
which, when inserted back into the Lagrangian (\ref{lagmr}), shows that the
$\mathcal{O}(M^2_S)$-terms cancel up to a constant.
Furthermore, the solution starts
at $\mathcal{O}(h^2)$ with coefficients
that are functions of $s_\alpha$, $c_\alpha$, $s_\beta$ and $c_\beta$.
Note that the combination
\begin{equation}                                                            
  \frac{s_\alpha c_\alpha^2}{s_\beta} + \frac{s_\alpha^2 c_\alpha}{c_\beta}
  = - c_{\beta-\alpha} [1 - 2 c_{\beta-\alpha}^2 +
  2 c_{\beta-\alpha} s_{\beta-\alpha} \cot{(2\beta)}]
\end{equation}
vanishes in the alignment limit.
Then, the square root in \eqref{h0exact} reduces to 1,
which gives $H_0(h)=0$. As a result, all tree level nondecoupling
effects vanish in the alignment limit.

\section{\boldmath One-loop matching of $h^n X_{\mu\nu}X^{\mu\nu}$
to all orders in $n$}
\label{sec:ynxx}

When calculating the one-loop EFT contributions of the form
$h^n X_{\mu\nu} X^{\mu\nu}$, we noted that, since $\hat{Y} = \mathcal{O}(M_H^2)$,
the series does not converge. Therefore, in order to calculate the full
Higgs function associated with the operator $X_{\mu\nu} X^{\mu\nu}$, we need
all coefficients $C_n$ of the expression
\begin{equation} \label{Coeffdef}
  \mathcal{L}_{\text{eff}} \supset \sum_{n=1}^\infty C_n \langle \hat{Y}^n
  \hat{X}_{\mu\nu} \hat{X}^{\mu\nu} \rangle
\end{equation}
In writing the above, we made essential use of the fact that
$\hat{Y} \propto\boldsymbol{1}$ and thus commutes with $\hat{X}^{\mu\nu}$.
This is, however, a special case. In general, $\hat{Y}$ does not
commute, giving more possible operator structures for each $n$.

To derive an all-orders result for the $C_n$, we start from the general
expression for the one-loop effective Lagrangian given in (\ref{leffp}).
We now use a slightly adapted form
of a trick explained in the Appendix of \cite{Bilenky:1993bt}:
We evaluate expression (\ref{leffp}) in the special configuration
$\partial_\mu \hat{X}_\nu = \partial_\mu \hat{Y}=0$,
allowing us to drop all derivatives. In this case
\begin{equation}
  D_\mu \hat{G} \rightarrow [\hat{X}_\mu, \hat{G}] \, , \qquad
  \hat{X}_{\mu \nu} \rightarrow [\hat{X}_\mu, \hat{X}_\nu] 
\end{equation}
where $\hat{G}$ is any matrix valued function of $\hat{Y}$ and $\hat{X}_\mu$.
In the final expressions, we can express everything through $D_\mu$ and
$\hat{X}_{\mu\nu}$, regaining the general result.

In our special case, we are only interested in the terms of the form
$\hat{Y}^n \hat{X}_{\mu \nu} \hat{X}^{\mu \nu}$, which contribute to the
$h^n \gamma \gamma$ nondecoupling effects. 
Setting $[\hat{X}_\mu,\hat{Y}] = 0$ automatically removes
all terms of the form $D_\mu \hat{Y}$.

In this way, it is easy to evaluate all terms from (\ref{leffp})
with 4 derivatives (4 factors of $X_\mu$) and $n$ factors of $Y$,
which reduce to the terms of interest due to the formal gauge invariance
of the functional integral. Finally we obtain
\begin{align}\label{leffynxx}
  \mathcal{L}_{\text{eff}} =
  \frac{1}{32\pi^2} \sum_{n=1}^\infty \frac{(-1)^n}{12 n M_H^{2n}}
   \langle \hat{Y}^n \hat{X}_{\mu\nu} \hat{X}^{\mu\nu} \rangle
  =\frac{X_{\mu\nu} X^{\mu\nu}}{192\pi^2}\ln\left( 1+\frac{Y(h)}{M^2_H}\right)
\end{align}
The first two terms in the sum over $n$ agree with those given
in~\cite{Fuentes-Martin:2016uol}
and the third and fourth terms agree with those given in
\cite{Banerjee:2023iiv}, in the special case that $\hat{Y}$ and
$\hat{X}_{\mu\nu}$ commute. 
The second expression in (\ref{leffynxx}) represents the full Higgs function
after taking
\begin{align}
  Y(h)=-d^{(0)}_5 h - d^{(0)}_7 H_0(h) - z^{(0)}_6 h^2
  - z^{(0)}_7 h H_0(h) - z^{(0)}_8 H_0(h)^2
\end{align}
and expressing $h$ in terms of the canonically normalized
Higgs field $\tilde h$, $h=h(\tilde h)$, through the inverse of
the function $\tilde h(h)$ defined in (\ref{htilde}).
Here $d^{(0)}_5$ is the ${\cal O}(M^2_S)$-part of
$d_5$, and similarly for the other coefficients,
and $H_0(h)$ is the function derived in Appendix \ref{sec:exh0}.

In the alignment limit $c_{\beta-\alpha}=0$, the Lagrangian
in (\ref{leffynxx}) can be given in closed form to all orders in $h$.
In this case, $H_0(h)$ vanishes and $Y$
in (\ref{xazy}) reduces to $Y/M^2_H = 2 h/v + (h/v)^2$. Then
(\ref{leffynxx}) takes the form of (\ref{lhff})
with the function $F_X(h)$ given by (\ref{fxfull}).

\section{Parameters of the 2HDM potential}
\label{sec:potpar}

The full scalar potential is given in (\ref{vdizi}).
In this section we give all coefficients of the potential
in terms of the input parameters
\begin{align}\label{inputpar}
  v, \quad m_h, \quad M_0, \quad M_H, \quad M_A, \quad \overline{m},
  \quad t_{\be}, \quad c_{\beta - \alpha}
\end{align}

The cubic couplings read

\begin{align}
  v \, d_1 &= c_{\beta - \alpha}^2 \lp \overline{m}^2 - m_h^2 \rp
             \lp s_{\beta - \alpha} + c_{\beta-\alpha} \cot(2\be)
             \rp  - \frac{m_h^2}{2} s_{\beta-\alpha} \\
  v \, d_2 &=  \frac{c_{\beta - \alpha}}{2} \left[ \left( 2m_h^2 + M_0^2 -
       3 \overline{m}^2 \right) \left( 1 - 2 c_{\be - \al}^2 +
    2 s_{\be-\al} c_{\be-\al} \cot (2\beta) \right) - \overline{m}^2 \right] \\
  v \, d_3 &=  -\frac{s_{\beta - \alpha}}{2} \left[ \left( 2 M_0^2 + m_h^2 -
    3 \overline{m}^2 \right) \left( 1 - 2 c_{\be - \al}^2 +
    2 s_{\be-\al} c_{\be-\al} \cot (2\beta) \right) + \overline{m}^2 \right] \\
  v \, d_4 &= s_{\beta - \alpha}^2 \left( \overline{m}^2 - M_0^2 \right)
             \left( c_{\beta - \alpha} - s_{\beta-\alpha} \cot(2\be)   \right)  -
             \frac{M_0^2}{2} c_{\beta-\alpha} \\
  v \, d_5 &=  2 s_{\beta - \alpha} \left( \overline{m}^2 - M_H^2 -
             \frac{m_h^2}{2} \right) + 2 c_{\beta-\alpha} \cot(2\beta)
             \left( \overline{m}^2 - m_h^2 \right) \\
  v \, d_6 &=  s_{\beta - \alpha} \left( \overline{m}^2 - M_A^2 -
             \frac{m_h^2}{2}\right) + c_{\beta-\alpha} \cot(2\beta)
             \left( \overline{m}^2 - m_h^2 \right) \\
  v \, d_7 &= 2 c_{\beta - \alpha} \left( \overline{m}^2 - M_H^2 -
             \frac{M_0^2}{2}\right) + 2 s_{\beta-\alpha} \cot(2\beta)
             \left( M_0^2 - \overline{m}^2\right) \\
  v \, d_8 &= c_{\beta - \alpha} \left( \overline{m}^2 - M_A^2 -
             \frac{M_0^2}{2}\right) + s_{\beta-\alpha} \cot(2\beta)
             \left( M_0^2 - \overline{m}^2\right)
\end{align}
and the quartic couplings are given by
\begin{align}
  v^2 z_1 =& -\frac{m_h^2}{8} + \frac{c_{\beta-\alpha}^2}{8}
             \left[ 4 s_{\be-\al}^2 \overline{m}^2 +(-3+4c_{\be-\al}^4) m_h^2 -
          \lp 1- 2c_{\be-\al}^2 \rp^2 M_0^2 \right. \nonumber \\
       &+ 4 c_{\be-\al} s_{\be-\al} \cot(2\be) \left( 2 \overline{m}^2 -
         \left( 1 + 2 c_{\be-\al}^2\right) m_h^2 -
         \left( 1 - 2 c_{\be-\al}^2\right) M_0^2 \right) \nonumber \\
           & \left. + 4 c_{\be-\al}^2 \cot^2(2\be) \left( \overline{m}^2 -
             c_{\be-\al}^2 m_h^2 - s_{\be-\al}^2 M_0^2 \right) \right] \\
  v^2 z_2 =& \frac{s_{\be-\al}c_{\be-\al}}{2} \left( 1 - 2c_{\be-\al}^2 \right)
             \left[ m_h^2 + M_0^2 - 2 c_{\be-\al}^2
        \left( M_0^2 -m_h^2\right) - 2 \overline{m}^2 \right] \nonumber \\
        &+ c_{\beta-\al}^2 \cot(2\be) \left[ m_h^2 \left( 1 + 2 c_{\be-\al}^2 -
          4 c_{\be-\al}^4 \right) + 2 M_0^2 \left( 1 - 3 c_{\be-\al}^2 +
          2 c_{\be-\al}^4 \right)  \right.  \nonumber\\         
  & + \left. \overline{m}^2 \left( -3 + 4 c_{\be-\al}^2 \right) \right] 
      + 2 c_{\be-\al}^3 s_{\be-\al} \cot^2(2\be) \left[ M_0^2 -
      \overline{m}^2 - c_{\be-\al}^2 \left( M_0^2 -m_h^2 \right) \right] \\
  v^2 z_3 =& \frac{1}{4} \left[ \left( 2 - 12 c_{\be-\al}^2  +
             12 c_{\be-\al}^4 \right) \overline{m}^2  \right. \nonumber \\
  & +\left( 1 - 2 c_{\be-\al}^2\right) \left( \left( -1 - 3 c_{\be -\al}^2 +
    6 c_{\be -\al}^4 \right) m_h^2 + \left( -2 + 9 c_{\be -\al}^2 -
    6 c_{\be -\al}^4 \right)  M_0^2 \right) \nonumber \\
       &+ 2 c_{\be-\al} s_{\be-\al} \cot(2\be) \left( \left( 6 -
  12 c_{\be -\al}^2   \right)  \overline{m}^2  +\left( -1 - 6 c_{\be -\al}^2 +
         12 c_{\be -\al}^4 \right) m_h^2 \right.   \nonumber \\
           &\left. + \left( -5 + 18  c_{\be -\al}^2 - 12 c_{\be -\al}^4 \right)
             M_0^2 \right) \nonumber \\
  & \left. + 12 c_{\be-\al}^2  s_{\be-\al}^2 \cot^2(2\be) \left( \overline{m}^2 -
    s_{\be-\al}^2 M_0^2 - c_{\be-\al}^2 m_h^2 \right)  \right] \\
  v^2 z_4 =& \frac{s_{\be-\al}c_{\be-\al}}{2} \left( 1 - 2c_{\be-\al}^2 \right)
       \left[ m_h^2 - 3 M_0^2 + 2 c_{\be-\al}^2 \left( M_0^2 -m_h^2 \right) +
             2 \overline{m}^2 \right] \nonumber \\
           &+ s_{\beta-\al}^2 \cot(2\be) \left[ m_h^2 \left( 2 c_{\be-\al}^2 -
             4 c_{\be-\al}^4 \right) +  M_0^2 \left( 1 - 6 c_{\be-\al}^2 +
             4 c_{\be-\al}^4 \right) \right. \nonumber\\
      &+ \left. \overline{m}^2 \left( -1 + 4 c_{\be-\al}^2 \right) \right]
   + 2 c_{\be-\al} s_{\be-\al}^3 \cot^2(2\be) \left[ M_0^2 - \overline{m}^2 -
     c_{\be-\al}^2 \left( M_0^2 -m_h^2 \right)  \right] \\
  v^2 z_5 =& \frac{1}{8} \left[ 4 s_{\be-\al}^2 c_{\be-\al}^2 \overline{m}^2 -
             s_{\be-\al}^2 \left( 1- 2c_{\be-\al}^2\right)^2 m_h^2 - c_{\be-\al}^2
             \left( 3- 2c_{\be-\al}^2 \right)^2 M_0^2 \right. \nonumber \\
   &+ 4 c_{\be-\al} s_{\be-\al}^3 \cot(2\be) \left( - 2 \overline{m}^2  +
     \left( -1 + 2 c_{\be-\al}^2\right) m_h^2 +\left( 3 - 2c_{\be-\al}^2 \right)
     M_0^2 \right) \nonumber \\
       & \left. + 4 s_{\be-\al}^4 \cot^2(2\be) \left( \overline{m}^2 -M_0^2 +
      c_{\be-\al}^2 \left( M_0^2- m_h^2 \right) \right)  \right] 
    \end{align}
    \begin{align}
      v^2 z_6 =& \frac{1}{2} \left[ 2 s_{\be-\al}^2
        \left( \overline{m}^2 - M_H^2 \right) - m_h^2 + c_{\be-\al}^2
  \left( 1 - 2 c_{\be-\al}^2 \right) \left( M_0^2 - m_h^2 \right) \right.
                 \nonumber \\
 &+ 2 c_{\be-\al} s_{\be-\al} \cot(2\be) \left( 2 \overline{m}^2 - m_h^2 -
  M_0^2 + 3 c_{\be-\al}^2 \left( M_0^2 - m_h^2 \right) \right) \nonumber \\
  & \left. + 4 c_{\be-\al}^2 \cot^2(2\be) \left( \overline{m}^2 - s_{\be-\al}^2
        M_0^2 - c_{\be-\al}^2 m_h^2 \right)  \right] \\
  v^2 z_{7} =&  c_{\be-\al} s_{\be-\al} \left( 2 \overline{m}^2 - 2 M_H^2 -
  \left( 1- 2 c_{\be-\al}^2 \right) \left( M_0^2 - m_h^2 \right) \right)
               \nonumber \\
 + &2 \cot(2\be) \left( M_0^2 -\overline{m}^2 +2 c_{\be-\al}^2 \overline{m}^2 +
   c_{\be-\al}^2 M_0^2 \left( -4 + 3 c_{\be-\al}^2 \right) \right. \nonumber\\
  + & c_{\be-\al}^2 \left. m_h^2 \left( 2 - 3 c_{\be-\al}^2 \right) \right)    
  + 4 c_{\be-\al} s_{\be-\al} \cot^2(2\be) \left(  M_0^2 - \overline{m}^2 -
       c_{\be-\al}^2 \left( M_0^2 -m_h^2 \right) \right)  \\
    v^2 z_{8} =& \frac{1}{2} \left[ -m_h^2 + c_{\be-\al}^2
  \left( 2 \left( \overline{m}^2 - M_H^2 \right) + \left( M_0^2 - m_h^2 \right)
    \left( -3 + 2 c_{\be-\al}^2\right) \right) \right. \nonumber \\
   &+ 2 c_{\be-\al} s_{\be-\al} \cot(2\be) \lp - 2 \overline{m}^2 - 2m_h^2 +
      4M_0^2 - 3 c_{\beta-\alpha}^2 \lp M_0^2-m_h^2  \rp \rp \nonumber \\
     & \left. + 4 s_{\be-\al}^2 \cot^2(2\be) \left( \overline{m}^2 - M_0^2 +
       c_{\be-\al}^2 \left( M_0^2 -m_h^2 \right) \right) \right] \\
   v^2 z_{9} =& \frac{1}{2} \left[ -s_{\beta-\alpha}^2 m_h^2 - c_{\beta-\alpha}^2
  M_0^2 + 4 \cot(2\be) c_{\be-\al} s_{\be-\al} \left( M_0^2 -m_h^2 \right)
                \right. \nonumber \\
   & \left. + 4 \cot^2(2\be) \left( \overline{m}^2 - M_0^2 + c_{\be-\al}^2
       \left( M_0^2 -m_h^2 \right) \right) \right] \\
   v^2 z_{10} =& \frac{1}{4} \left[ 2 s_{\beta-\alpha}^2 \left( \overline{m}^2 -
    M_A^2 \right) - m_h^2 \left( 1 + c_{\be-\al}^2 - 2 c_{\be-\al}^4 \right) +
   c_{\be-\al}^2 M_0^2 \left( 1 - 2 c_{\be-\al}^2 \right)  \right. \nonumber \\
   & + 2 c_{\be-\al} s_{\be-\al} \cot(2\be) \left( 2 \overline{m}^2 - m_h^2 -
   M_0^2 + 3 c_{\beta-\alpha}^2 \left( M_0^2-m_h^2  \right) \right) \nonumber \\
        &\left. + 4 c_{\be-\al}^2 \cot^2(2\be)  \left( \overline{m}^2 -
          s_{\be-\al}^2 M_0^2 - c_{\be-\al}^2 m_h^2 \right) \right] \\
  v^2 z_{11} =& \frac{1}{2} \left[ c_{\be-\al} s_{\be-\al}
       \left( 2 \overline{m}^2 - 2 M_A^2 + \left( 2 c_{\be-\al}^2 -1 \right)
            \left( M_0^2 - m_h^2 \right) \right)  \right. \nonumber \\
 + &2 \cot(2\be) \left( M_0^2 - \overline{m}^2 +2 c_{\be-\al}^2 \overline{m}^2 +
   c_{\be-\al}^2 M_0^2 \left( -4 + 3 c_{\be-\al}^2 \right) + c_{\be-\al}^2 m_h^2
   \left( 2 - 3 c_{\be-\al}^2 \right) \right) \nonumber \\
    & \left. + 4 c_{\be-\al} s_{\be-\al} \cot^2(2\be) \left(  M_0^2 -
\overline{m}^2 - c_{\be-\al}^2 \left( M_0^2 -m_h^2 \right) \right) \right] \\ 
    v^2 z_{12} =& \frac{1}{4} \left[ -m_h^2 + c_{\be-\al}^2
      \left( 2 \overline{m}^2 + 3 m_h^2 - 2M_A^2 -3 M_0^2 \right) +
     2 c_{\be-\al}^4 \left( M_0^2 -m_h^2\right) \right. \nonumber \\
      &+ 2 c_{\be-\al} s_{\be-\al} \cot(2\be) \left( 4 M_0^2 - 2m_h^2 -2
     \overline{m}^2 - 3 c_{\beta-\alpha}^2 \left( M_0^2-m_h^2  \right) \right)
        \nonumber \\
    & \left.  + 4 s_{\be-\al}^2 \cot^2(2\be) \left( \overline{m}^2 - M_0^2 +
      c_{\be-\al}^2 \left( M_0^2 -m_h^2 \right) \right) \right] \\
  v^2 z_{13} =& 4 v^2 z_{14} = \frac{1}{2} \left[ -s_{\beta-\alpha}^2 m_h^2 -
       c_{\beta-\alpha}^2 M_0^2 + 4 \cot(2\be) c_{\be-\al} s_{\be-\al}
                \left( M_0^2 -m_h^2 \right)   \right. \nonumber \\
    & \left. + 4 \cot^2(2\be) \left( \overline{m}^2 - M_0^2 + c_{\be-\al}^2
                 \left( M_0^2 -m_h^2 \right) \right) \right] 
\end{align}

In the alignment limit ($c_{\beta-\alpha} \to 0$) the coupling constants
simplify to
\begin{align}
  v d_1 &= - \frac{m_h^2}{2}, \quad v d_2 = 0, \quad
    v d_3= \overline{m}^2 - M_0^2 - \frac{m_h^2}{2}  \nonumber \\
  v d_5 &= 2\overline{m}^2 - 2 M_H^2 - m_h^2, \quad
      v d_6 = \overline{m}^2 - M_A^2 - \frac{m_h^2}{2}, \nonumber \\
  d_4 &= \frac{1}{2} d_7 = d_8 =
        \cot(2\be) \frac{\left( M_0^2 - \overline{m}^2 \right)}{v}
\end{align}
\begin{align}
  v^2 z_1 &= - \frac{m_h^2}{8}, \quad v^2 z_2 = 0, \quad v^2 z_3 =
    \frac{1}{2} \left( \overline{m}^2 - M_0^2 - \frac{m_h^2}{2} \right),
 \quad  v^2  z_6 = \overline{m}^2 - M_H^2 - \frac{m_h^2}{2}  \nonumber \\
  z_4 &= \frac{z_7}{2} = z_{11} =
   \cot(2\be) \frac{\left( M_0^2 - \overline{m}^2 \right)}{v^2} , \quad
    v^2 z_{10} = -\frac{1}{2} \left( M_A^2 - \overline{m}^2 \right) -
        \frac{m_h^2}{4} \nonumber \\
  4 z_5 &= z_8 =  z_9 = 2z_{12} = z_{13} = 4 z_{14} = - \frac{m_h^2}{2 v^2} -
        2 \cot^2(2\be) \frac{\left( M_0^2 - \overline{m}^2 \right)}{v^2}
\end{align}

It is also useful to express the parameters of the potential
in the original form of (\ref{vs1s2}) in terms of the
physical parameters in (\ref{inputpar}). The relations read
\begin{align}
  m^2_{11} &= s^2_\beta  \overline{m}^2
       - \frac{1}{2} (c^2_\alpha M^2_0 + s^2_\alpha m^2_h)
     - \frac{s^2_\beta}{2} \frac{s_{2\alpha}}{s_{2\beta}} (M^2_0 - m^2_h) \\
  m^2_{22} &= c^2_\beta  \overline{m}^2
       - \frac{1}{2} (s^2_\alpha M^2_0 + c^2_\alpha m^2_h)
       - \frac{c^2_\beta}{2} \frac{s_{2\alpha}}{s_{2\beta}} (M^2_0 - m^2_h) \\
  \lambda_1 &= \frac{1}{c^2_\beta v^2}
              (c^2_\alpha M^2_0 + s^2_\alpha m^2_h - s^2_\beta \overline{m}^2) \\
 \lambda_2 &= \frac{1}{s^2_\beta v^2}
              (s^2_\alpha M^2_0 + c^2_\alpha m^2_h - c^2_\beta \overline{m}^2) \\
  \lambda_3 &=\frac{1}{v^2}\left(2 M^2_H -\overline{m}^2 +
              \frac{s_{2\alpha}}{s_{2\beta}} (M^2_0 - m^2_h)\right) \\
  \lambda_4 &= \frac{1}{v^2} (M^2_A - 2 M^2_H + \overline{m}^2) \\
  \lambda_5 &= \frac{1}{v^2} (\overline{m}^2 - M^2_A)
\end{align}  
The absence of a decoupling limit for $\overline{m}=0$ is obvious
from these formulas.



\begin{thebibliography}{99}

\bibitem{Feruglio:1992wf} 
  F.~Feruglio,
  Int.\ J.\ Mod.\ Phys.\ A {\bf 08} (1993) 4937 
  [hep-ph/9301281].

\bibitem{Bagger:1993zf} 
  J.~Bagger {\it et. al.}, 
  Phys.\ Rev.\ D {\bf 49} (1994) 1246 
  [hep-ph/9306256].

\bibitem{Koulovassilopoulos:1993pw} 
  V.~Koulovassilopoulos and R.~S.~Chivukula,
  Phys.\ Rev.\ D {\bf 50} (1994) 3218 
  [hep-ph/9312317].

\bibitem{Burgess:1999ha} 
  C.~P.~Burgess, J.~Matias and M.~Pospelov,
  Int.\ J.\ Mod.\ Phys.\ A {\bf 17} (2002) 1841
  [hep-ph/9912459].

\bibitem{Wang:2006im}
L.~M.~Wang and Q.~Wang,
[arXiv:hep-ph/0605104 [hep-ph]].
  
\bibitem{Grinstein:2007iv} 
  B.~Grinstein and M.~Trott,
  Phys.\ Rev.\ D {\bf 76} (2007) 073002 
  [arXiv:0704.1505 [hep-ph]].

\bibitem{Contino:2010mh} 
  R.~Contino, C.~Grojean, M.~Moretti, F.~Piccinini and R.~Rattazzi,
  JHEP {\bf 1005} (2010) 089 
  [arXiv:1002.1011 [hep-ph]].

\bibitem{Contino:2010rs} 
  R.~Contino,
  arXiv:1005.4269 [hep-ph].

\bibitem{Buchalla:2012qq}
G.~Buchalla and O.~Cat\`a,
JHEP \textbf{07} (2012) 101
[arXiv:1203.6510 [hep-ph]].
  
\bibitem{Alonso:2012px} 
  R.~Alonso, M.~B.~Gavela, L.~Merlo, S.~Rigolin and J.~Yepes,
  Phys.\ Lett.\ B {\bf 722} (2013) 330 
  Erratum: [Phys.\ Lett.\ B {\bf 726} (2013) 926]
  [arXiv:1212.3305 [hep-ph]].

\bibitem{Alonso:2012pz} 
  R.~Alonso, M.~B.~Gavela, L.~Merlo, S.~Rigolin and J.~Yepes,
  Phys.\ Rev.\ D {\bf 87} (2013) no. 5, 055019
  [arXiv:1212.3307 [hep-ph]].

\bibitem{Buchalla:2013rka} 
  G.~Buchalla, O.~Cat\`a and C.~Krause,
  Nucl.\ Phys.\ B {\bf 880} (2014) 552 
  Erratum: [Nucl.\ Phys.\ B {\bf 913} (2016) 475]
  [arXiv:1307.5017 [hep-ph]].

\bibitem{Delgado:2014jda}
R.~L.~Delgado, A.~Dobado, M.~J.~Herrero and J.~J.~Sanz-Cillero,
JHEP \textbf{07} (2014) 149
[arXiv:1404.2866 [hep-ph]].
  
\bibitem{Buchalla:2015qju}
G.~Buchalla, O.~Cat\`a, A.~Celis and C.~Krause,
Eur. Phys. J. C \textbf{76} (2016) no.5, 233
[arXiv:1511.00988 [hep-ph]].

\bibitem{Alonso:2016oah}
R.~Alonso, E.~E.~Jenkins and A.~V.~Manohar,
JHEP \textbf{08} (2016), 101
[arXiv:1605.03602 [hep-ph]].

\bibitem{Cohen:2020xca}
T.~Cohen, N.~Craig, X.~Lu and D.~Sutherland,
JHEP \textbf{03} (2021) 237
[arXiv:2008.08597 [hep-ph]].

\bibitem{Gomez-Ambrosio:2022qsi}
R.~G\'omez-Ambrosio, F.~J.~Llanes-Estrada, A.~Salas-Bern\'ardez and
J.~J.~Sanz-Cillero,
Phys. Rev. D \textbf{106} (2022) no.5, 053004
[arXiv:2204.01763 [hep-ph]].

\bibitem{Gomez-Ambrosio:2022why}
R.~G\'omez-Ambrosio, F.~J.~Llanes-Estrada, A.~Salas-Bern\'ardez and
J.~J.~Sanz-Cillero,
Commun. Theor. Phys. \textbf{75} (2023) no.9, 095202
[arXiv:2207.09848 [hep-ph]];
EPJ Web Conf. \textbf{274} (2022) 08013
[arXiv:2211.09605 [hep-ph]].

\bibitem{Delgado:2023ynh}
R.~L.~Delgado, R.~G\'omez-Ambrosio, J.~Mart\'\i{}nez-Mart\'\i{}n,
A.~Salas-Bern\'ardez and J.~J.~Sanz-Cillero,
[arXiv:2311.04280 [hep-ph]].

\bibitem{Buchalla:2015wfa}
G.~Buchalla, O.~Cat\`a, A.~Celis and C.~Krause,
Phys. Lett. B \textbf{750} (2015) 298-301
[arXiv:1504.01707 [hep-ph]].

\bibitem{Branco:2011iw}
G.~C.~Branco, P.~M.~Ferreira, L.~Lavoura, M.~N.~Rebelo,
M.~Sher and J.~P.~Silva,
Phys. Rept. \textbf{516} (2012) 1
[arXiv:1106.0034 [hep-ph]].

\bibitem{Dmytriiev:2022asw}
M.~Dmytriiev and V.~Skalozub,
Journal of Physics and Electronics \textbf{29} (2022) no.2, 8
[arXiv:2206.07770 [hep-ph]].

\bibitem{Banta:2023prj}
I.~Banta, T.~Cohen, N.~Craig, X.~Lu and D.~Sutherland,
JHEP \textbf{06} (2023), 150
[arXiv:2304.09884 [hep-ph]].

\bibitem{Dawson:2023ebe}
S.~Dawson, D.~Fontes, C.~Quezada-Calonge and J.~J.~Sanz-Cillero,
Phys. Rev. D \textbf{108} (2023) no.5, 055034
[arXiv:2305.07689 [hep-ph]].

\bibitem{Arco:2023sac}
F.~Arco, D.~Domenech, M.~J.~Herrero and R.~A.~Morales,
[arXiv:2307.15693 [hep-ph]].

\bibitem{Ciafaloni:1996ur}
P.~Ciafaloni and D.~Espriu,
Phys. Rev. D \textbf{56} (1997) 1752
[arXiv:hep-ph/9612383 [hep-ph]].

\bibitem{Buchalla:2016bse}
G.~Buchalla, O.~Cat\`a, A.~Celis and C.~Krause,
Nucl. Phys. B \textbf{917} (2017) 209
[arXiv:1608.03564 [hep-ph]].

\bibitem{Dittmaier:2022ivi}
S.~Dittmaier and H.~Rzehak,
JHEP \textbf{08} (2022) 245
[arXiv:2206.01479 [hep-ph]].

\bibitem{Fuentes-Martin:2016uol}
J.~Fuentes-Martin, J.~Portoles and P.~Ruiz-Femenia,
JHEP \textbf{09} (2016) 156
[arXiv:1607.02142 [hep-ph]].

\bibitem{Shifman:1979eb}
M.~A.~Shifman, A.~I.~Vainshtein, M.~B.~Voloshin and V.~I.~Zakharov,
Sov. J. Nucl. Phys. \textbf{30} (1979) 711



\bibitem{Gunion:2002zf}
J.~F.~Gunion and H.~E.~Haber,
Phys. Rev. D \textbf{67} (2003) 075019
[arXiv:hep-ph/0207010 [hep-ph]].

\bibitem{Arhrib:2003vip}
A.~Arhrib, M.~Capdequi Peyranere, W.~Hollik and S.~Penaranda,
Phys. Lett. B \textbf{579} (2004) 361
[arXiv:hep-ph/0307391 [hep-ph]].

\bibitem{Nebot:2019qvr}
M.~Nebot,
Phys. Rev. D \textbf{102} (2020) no.11, 115002
[arXiv:1911.02266 [hep-ph]].

\bibitem{Faro:2020qyp}
F.~Faro, J.~C.~Romao and J.~P.~Silva,
Eur. Phys. J. C \textbf{80} (2020) no.7, 635
[arXiv:2002.10518 [hep-ph]].


\bibitem{Huffel:1980sk}
H.~H{\"u}ffel and G.~P{\'o}csik,
Z. Phys. C \textbf{8} (1981) 13

\bibitem{Casalbuoni:1987cz}
R.~Casalbuoni, D.~Dominici, F.~Feruglio and R.~Gatto,
Nucl. Phys. B \textbf{299} (1988) 117

\bibitem{Maalampi:1991fb}
J.~Maalampi, J.~Sirkka and I.~Vilja,
Phys. Lett. B \textbf{265} (1991) 371

\bibitem{Kanemura:1993hm}
S.~Kanemura, T.~Kubota and E.~Takasugi,
Phys. Lett. B \textbf{313} (1993) 155
[arXiv:hep-ph/9303263 [hep-ph]].

\bibitem{Akeroyd:2000wc}
A.~G.~Akeroyd, A.~Arhrib and E.~M.~Naimi,
Phys. Lett. B \textbf{490} (2000) 119
[arXiv:hep-ph/0006035 [hep-ph]].

\bibitem{Horejsi:2005da}
J.~Horejsi and M.~Kladiva,
Eur. Phys. J. C \textbf{46} (2006) 81
[arXiv:hep-ph/0510154 [hep-ph]].

\bibitem{Goodsell:2018tti}
M.~D.~Goodsell and F.~Staub,
Eur. Phys. J. C \textbf{78} (2018) no.8, 649
[arXiv:1805.07306 [hep-ph]].

\bibitem{Cacchio:2016qyh}
V.~Cacchio, D.~Chowdhury, O.~Eberhardt and C.~W.~Murphy,
JHEP \textbf{11} (2016) 026
[arXiv:1609.01290 [hep-ph]].

\bibitem{Grinstein:2015rtl}
B.~Grinstein, C.~W.~Murphy and P.~Uttayarat,
JHEP \textbf{06} (2016) 070
[arXiv:1512.04567 [hep-ph]].

\bibitem{Bahl:2022jnx}
H.~Bahl, J.~Braathen and G.~Weiglein,
Phys. Rev. Lett. \textbf{129} (2022) 231802
[arXiv:2202.03453 [hep-ph]].

\bibitem{Deshpande:1977rw}
N.~G.~Deshpande and E.~Ma,
Phys. Rev. D \textbf{18} (1978) 2574

\bibitem{Klimenko:1984qx}
K.~G.~Klimenko,
Theor. Math. Phys. \textbf{62} (1985) 58

\bibitem{Maniatis:2006fs}
M.~Maniatis, A.~von Manteuffel, O.~Nachtmann and F.~Nagel,
Eur. Phys. J. C \textbf{48} (2006) 805
[arXiv:hep-ph/0605184 [hep-ph]].

\bibitem{Ivanov:2006yq}
I.~P.~Ivanov,
Phys. Rev. D \textbf{75} (2007) 035001
[erratum: Phys. Rev. D \textbf{76} (2007) 039902]
[arXiv:hep-ph/0609018 [hep-ph]].

\bibitem{Ivanov:2015nea}
I.~P.~Ivanov and J.~P.~Silva,
Phys. Rev. D \textbf{92} (2015) no.5, 055017
[arXiv:1507.05100 [hep-ph]].

\bibitem{Appelquist:1974tg}
T.~Appelquist and J.~Carazzone,
Phys. Rev. D \textbf{11} (1975) 2856

\bibitem{Perez:1995dc}
M.~A.~Perez, J.~J.~Toscano and J.~Wudka,
Phys. Rev. D \textbf{52} (1995) 494
[arXiv:hep-ph/9506457 [hep-ph]].

\bibitem{Dawson:2022cmu}
S.~Dawson, D.~Fontes, S.~Homiller and M.~Sullivan,
Phys. Rev. D \textbf{106} (2022) no.5, 055012
[arXiv:2205.01561 [hep-ph]].


\bibitem{deBlas:2018tjm}
J.~de Blas, O.~Eberhardt and C.~Krause,
JHEP \textbf{07} (2018) 048
[arXiv:1803.00939 [hep-ph]].

\bibitem{Asner:2002aa}
D.~Asner {\it et al.},
[arXiv:hep-ph/0208219 [hep-ph]].

\bibitem{Chowdhury:2017aav}
D.~Chowdhury and O.~Eberhardt,
JHEP \textbf{05} (2018) 161
[arXiv:1711.02095 [hep-ph]].

\bibitem{Arco:2022xum}
F.~Arco, S.~Heinemeyer and M.~J.~Herrero,
Eur. Phys. J. C \textbf{82} (2022) no.6, 536
[arXiv:2203.12684 [hep-ph]].

\bibitem{Bhattacharyya:2014oka}
G.~Bhattacharyya and D.~Das,
Phys. Rev. D \textbf{91} (2015) 015005
[arXiv:1408.6133 [hep-ph]].



\bibitem{Dawson:2023oce}
S.~Dawson, D.~Fontes, C.~Quezada-Calonge and J.~J.~Sanz-Cillero,
[arXiv:2311.16897 [hep-ph]].


\bibitem{Bilenky:1993bt}
M.~S.~Bilenky and A.~Santamaria,
Nucl. Phys. B \textbf{420} (1994) 47
[arXiv:hep-ph/9310302 [hep-ph]].

\bibitem{Banerjee:2023iiv}
U.~Banerjee, J.~Chakrabortty, S.~U.~Rahaman and K.~Ramkumar,
[arXiv:2306.09103 [hep-ph]].

  
\end{thebibliography}
\end{document}